# 3D magnetic field configuration of small-scale reconnection events in the solar plasma atmosphere

T. Shimizu[1,2,a]

[1]*Institute of Space and Astronautical Science, Japan Aerospace Exploration Agency, 3-1-1 Yoshinodai, Chuo-ku, Sagamihara, Kanagawa, 252-5210, Japan*

[2]*Department of Earth and Planetary Science, The University of Tokyo, 7-3-1 Hongo, Bunkyo-ku, Tokyo 113-0033 Japan*

The outer solar atmosphere, i.e., the corona and the chromosphere, is replete with small energy-release events, which are accompanied by transient brightening and jet-like ejections. These events are considered to be magnetic reconnection events in the solar plasma, and their dynamics have been studied using recent advanced observations from the *Hinode* spacecraft and other observatories in space and on the ground. These events occur at different locations in the solar atmosphere, and vary in their morphology and amount of the released energy. The magnetic field configurations of these reconnection events are inferred based on observations of magnetic fields at the photospheric level. Observations suggest that these magnetic configurations can be classified into two groups. In the first group, two anti-parallel magnetic fields reconnect to each other, yielding a 2D emerging flux configuration. In the second group, helical or twisted magnetic flux tubes are parallel or at a relative angle to each other. Reconnection can occur only between anti-parallel components of the magnetic flux tubes and may be referred to as component reconnection. The latter configuration type may be more important for the larger class of small-scale reconnection events. The two types of magnetic configurations can be compared to counter-helicity and co-helicity configurations, respectively, in laboratory plasma collision experiments.

## I. INTRODUCTION

Soft X-ray and extreme ultraviolet (EUV) imaging observations from space-borne observatories, such as *Yohkoh*, *SoHO*, *TRACE*, *Hinode*, and *SDO*, have revealed that microflares and jets are frequently produced in the solar corona. Recent high spatial resolution observations of the chromosphere by the *Hinode*[1] solar optical telescope (SOT) [2,3,4,5] have revealed that the chromosphere is also replete with transient brightening events and jet-like ejections. Magnetic reconnection has been assumed as a central player in converting the magnetic energy to kinetic and thermal energies on a short timescale (a few seconds to tens of minutes), resulting in the production of transient brightening events and dynamical jets. The central question associated with magnetic reconnection events is the mechanism of fast-timescale magnetic reconnection. Transient solar phenomena, especially flares, are not confined only to local reconnection regions, and the entire system of magnetic structures involving the reconnection regions changes its topology with eruption. Thus, the question is what role is assumed by local reconnections in large-scale topological changes. Moreover, trigger mechanism of the magnetic reconnections in the solar atmosphere is one of the greatest challenges for predicting the occurrence of solar flares. The entire system of magnetic structures gradually evolves in the solar atmosphere with inclusion of flux emergence and cancellation at the solar surface. In these conditions, magnetic reconnections are triggered locally, and small-scale reconnections self-organize into larger-scale ones.

---

[a] Electronic mail: shimizu@solar.isas.jaxa.jp

To address these fundamental questions related to magnetic reconnections, it is necessary to understand the overall 3D magnetic field configuration, including the reconnection regions. However, owing to its low emissivity and small spatial scale it is difficult to explore the plasma inside the local reconnection regions. With recent high spatial resolution observations it became possible to investigate the plasma in these local reconnection regions, especially in inflow and outflow regions of the reconnections[6,7]. Moreover, direct measurements of magnetic field vectors near the height of the reconnection are useful for inferring the structure of the 3D magnetic field formed in the corona and chromosphere. The inferred 3D magnetic field structure can help to determine the physical conditions and drivers for generation of non-steady reconnection events. The SOT spectro-polarimeter (SP)[7] records the full-polarization states of line profiles (Stokes spectral profiles) of two magnetically sensitive Fe I lines at 630.15 and 630.25 nm, and inversion analysis of the SP spectral data yields physical parameters at the solar surface (photosphere), including the magnetic field vectors and Doppler shifts.

In this paper, we review the recent observations of small-scale reconnection events, with particular emphasis on the 3D magnetic field structure involved in these events. Small-scale reconnection events, mainly based on the *Hinode* observations, are briefly reviewed in Section II. Their magnetic configurations are classified into two groups, which are discussed in Section III and summarized in Section IV.

## II. SMALL-SCALE RECONNECTION EVENTS

A variety of small-scale events accompanied by transient brightening and/or jet-like ejection have been reported up to date. These events have been differently termed. Figure 1 shows the distribution of small-scale energy release events as a function of the released energy (horizontal axis) and the height at which these events occurred (vertical axis). In the solar atmosphere, the plasma properties are strongly height-dependent. The corona is the rarefied outer atmosphere formed above the visible surface, i.e., the photosphere, and it is characterized by abnormally high temperature (above $10^6$ K) with the plasma density of the order of or under $10^9$ cm$^{-3}$. The coronal plasma is fully ionized and is predominantly controlled by magnetic pressure (plasma $\beta$ ~0.01). The electron mean free path is of the order of $10^4$ km. The plasma can be treated using a fluid formalism and its distribution functions are presumably close to Maxwellian. The chromosphere is the thin layer between the photosphere and the corona. The most common feature of the chromosphere is the presence of spicules, which are dynamic collimated jets with typical diameters of 200-500 km and length in the 2,000-10,000 km range. In the chromosphere, the temperature increases from its minimal value of nearly 4,000 K in the lower chromosphere to a few 10,000 K in the upper chromosphere, next to the transition layer of the corona. The density of hydrogen in the chromosphere changes from $10^{14}$ cm$^{-3}$ to $10^{11}$ cm$^{-3}$. The chromosphere hydrogen is partially ionized (degree of ionization in the 0.001-1 range) and the electron density is of the order of $10^{11}$ cm$^{-3}$. The electron mean free path is 10-100 cm and thus the plasma is collisional. Plasma β is high at the bottom of the chromosphere (about 500 km above the photosphere), where plasma behavior is affected by the gas dynamics; it becomes nearly unity at about 1,000 km above the photosphere and changes to the low β condition (~0.01) towards the upper chromosphere. In the following subsections, each of the small-scale energy release events will be briefly reviewed, with particular emphasis on the investigations of magnetic field configurations.

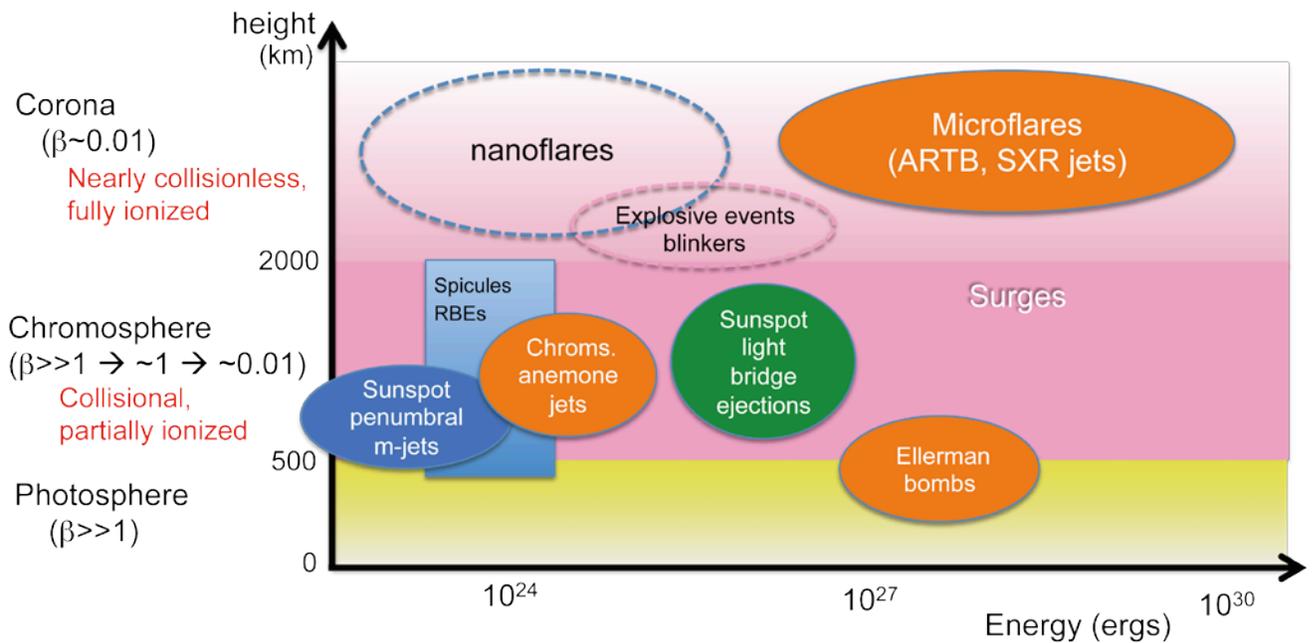

FIG. 1. Small-scale energy release events in the solar atmosphere. A variety of differently termed small-scale events accompanied by transient brightening events and jet-like ejections have been reported up to date. The released energy values on the abscissa show the typical values estimated for the different types of events.

**A. Coronal microflares, X-ray jets, and nanoflares**

Microflares correspond to energy release events in which the released energy is approximately 6 orders of magnitude smaller than the energy released by the largest flares, i.e., of the order of $10^{27}$ ergs. The *Yohkoh* soft X-ray observations revealed that active regions exhibit frequent transient brightening events of compact coronal loops[9]; these events are referred to as active-region transient brightenings (ARTBs) and sometimes as soft X-ray microflares. These events may be accompanied by X-ray jets[10]. The photospheric magnetograms, which yield the spatial distribution of the magnetic flux density on the solar surface, are used for deducing the overall magnetic configuration involved in the transient brightenings and for identifying the magnetic activities in the photosphere that are responsible for the onset of these brightening events. Opposite magnetic polarities encountering each other were found in the vicinity of about half of the studied transient brightenings[11][12]. Emerging magnetic flux, i.e., the magnetic flux appearing at the solar surface owing to the buoyancy-driven emergence process, was proposed as the dominant factor causing the encounters of opposite polarities; this suggests the emerging flux configuration in which a small magnetic bipole creates anti-parallel magnetic discontinuity with the pre-existing ambient magnetic field in the lower corona. Unipolar magnetic features moving out from the outer boundary of sunspots may also create anti-parallel magnetic discontinuity in the lower atmosphere via chance encounters with opposite-polarity magnetic islands. The emerging flux configuration model (Figure 2) predicts the appearance of a compact brightening source with the upward ejection of hot and cool plasma[13]. The model describes magnetic reconnections in a 2D domain. However, helical threads are sometimes observed to undergo untwisting ejection motion[14], leading to more complicated 3D pictures, such as blowout jets[15], in which a base emerging arch has enough shear or twist to erupt open. Shear or twist involved in the emerging flux bipole takes a role in creating twisted motion of ejections [16].

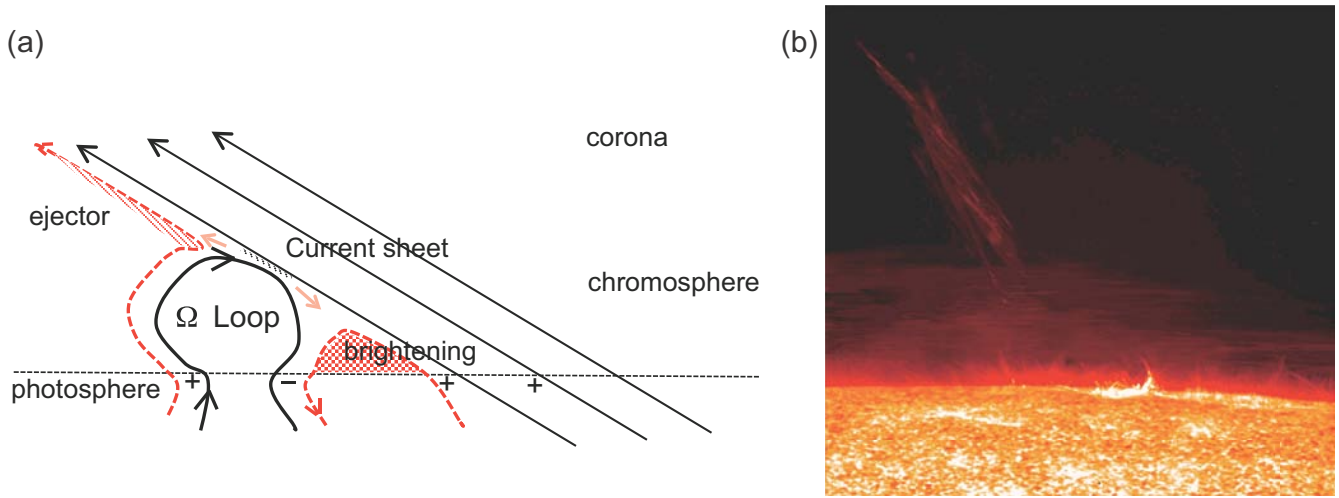

FIG. 2. (a) The emerging flux configuration model and (b) a large-scale class of chromospheric jets showing untwisting motions of ejectors. The image was taken using the Ca II H filter, sensitive to the chromospheric temperature (~$10^4$ K), by the SOT onboard *Hinode*. A compact transient brightening can be clearly seen near the jet footpoint on the solar surface.

Another key observation is that such magnetic encounters are not found in the vicinity of about half of the observed microflares. They sometimes consist of multiple coronal brightening loops[17] and their released energy tends to be larger than that of compact single-loop or point-like brightening events[9]. One example of such a microflare is shown in Figure 3. At least two or more coronal loop structures are almost simultaneously brightened. Two or more chromospheric brightening kernels can be clearly seen at each end of the brightening X-ray loops and they can be used to exactly identify the footpoints of the coronal brightening structures (Figure 3b). The several brightening kernels at one end of the coronal loops all have the same magnetic polarity (Figure 3d), implying that the main axis of the magnetic field is in the same direction in all of the brightening loops. At the sunspot umbra side, the footpoints of the brightening loops are located in the area in which the vertical electric current is enhanced (blue dominant in Figure 3f), suggesting that the magnetic field originating from the footpoints in the sunspot is twisted counterclockwise in the corona. When the twisted field is simplified by assuming two current-carrying loops (with helical field structure, Figure 4), the poloidal components of the helical structures are anti-parallel, which is preferable for magnetic reconnection. Moreover, the current-carrying loops may be braided in the corona, and the magnetic braids may reconnect, relax, and dissipate sufficient energy to the magnetic structures to reach the coronal temperatures. A highest spatial resolution EUV observation recently resolved magnetic braids in a coronal loop structure[18]. The idea of magnetic braids was originally proposed as a nanoflare hypothesis for heating the corona[19]; gas motion and dynamics at the solar surface (photosphere) move the footpoints of magnetic fields and form many discontinuities in the evolving magnetic fields in the corona.

The magnetized plasma may be more turbulent at lower temperature in the transition region compared with the higher coronal temperature. The transition region is the interface temperature layer between the hot ($>10^6$ K) corona and relatively cool ($\sim10^4$ K) chromosphere. High spectral resolution EUV spectra revealed explosive events[20], characterized by non-Gaussian enhancements of the red and blue wavelength wings in line profiles. These events were detected in the spectral lines formed between $2\times10^4$ K and $2\times10^5$ K and exhibited high Doppler shifts, ranging from +/- 50 km/s to +/- 250 km/s. A different type of EUV events were reported as UV blinkers[21], corresponding to tiny emission brightening events in the transition region plasma, lasting between 10 and 15 min, located in the supergranular cell corners. Blinkers may be the on-disk signature of chromospheric spicules[22]. A *SoHO* SUMER's UV observation demonstrated bi-directional jets in the transition region temperature, i.e., a blue-shifted enhancement was observed simultaneously with a red-shifted enhancement at a displacement along the slit, indicating the outflow jets that are theoretically expected from reconnection sites[23].

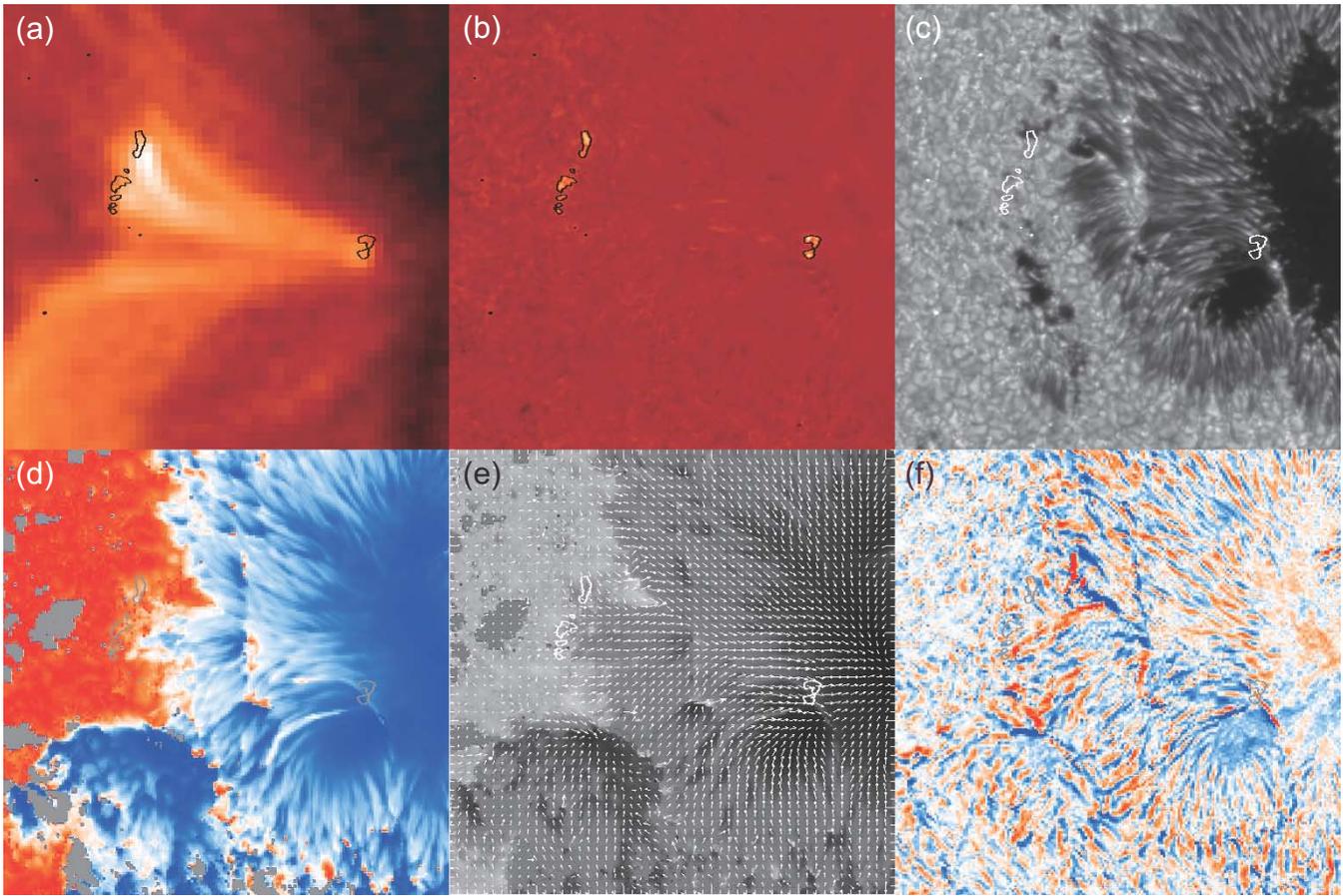

FIG. 3. A multiple-loop type microflare, observed by *Hinode*. (a) The soft X-ray image obtained by the *Hinode* X-Ray Telescope, (b) the footpoints of the brightening loops, identified by increased intensity in the chromospheric CaII H images, (c) a white light image, showing the sunspot structures, (d) the magnetic inclination (blue and red: vertical, white: horizontal), (e) the magnetic field vector, and (f) the vertical electrical currents in the photosphere.

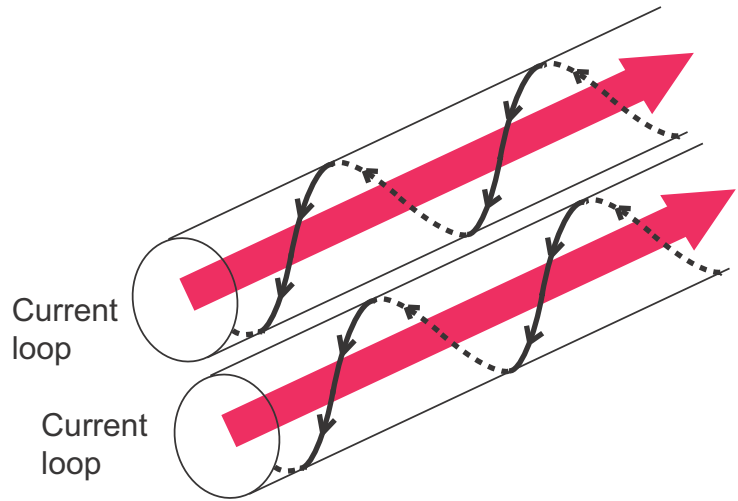

FIG. 4. Magnetic reconnection between two current-carrying loops, inferred as the magnetic field configuration of two brightening loops.

**B. Jet-like features in the chromosphere**

The *Hinode* SOT has been providing high spatial (0.2-0.3 arcsec) and high cadence (5 – 60 s) resolution images with a broadband filter covering the chromospheric Ca II H spectral line at 397 nm. One of the findings unexpected by many solar physicists was that the chromosphere is more dynamic than had ever been imagined. The SOT observations revealed various jet-like features, including ubiquitous anemone jets, light bridge ejections and penumbral microjets in sunspots, and spicules.

*1. Ubiquitous chromospheric anemone jets*

Small-scale chromospheric anemone jets are ubiquitously present outside the sunspots in the active regions[24], as seen in Figure 5. The jets typically have bright, inverted Y-shape structures, similar to the shapes of X-ray anemone jets in the corona, although they are smaller in size. The morphological similarity of chromospheric anemone jets and X-ray anemone jets suggests that they are produced as a result of the magnetic reconnection between a small-scale emerging (or moving) magnetic bipole and a pre-existing ambient field in the lower chromosphere. The jets are typically 2,000-5,000 km long and 150-300 km wide. A collimated plasma flow is launched from the top of a small and bright loop. The apparent speed is 5-20 km/s, close to the local Alfven speed in the lower chromosphere, and the lifetime is in the 100-500 s range[25]. Each collimated plasma flow exhibits temporal variations, and multiple plasma ejections may be involved in a collimated plasma flow; the height-time plot of jets reveals many individual ejections, demonstrating time-dependent as well as intermittent nature of magnetic reconnection in the chromosphere[26]. The Ca II H imaging observations can provide only the morphology and the apparent speed in the transversal plane measured from the line of sight direction; no physical parameters such as density and magnetic field strength can be deduced from the Ca II H images. Simultaneous acquisition of photospheric magnetograms shows that there may exist mixed polarity patches at the footpoints of chromospheric anemone jets, indicative of magnetic reconnection for the formation of chromospheric anemone jets. Assuming that the density of chromospheric jets is $10^{15}$ cm$^{-3}$, the overall thermal energy is of the order of $10^{25}$ ergs, slightly below the reduction in the magnetic flux energy of $10^{26}$-$10^{27}$ ergs owing to the magnetic cancellation observed in the mixed polarities[27].

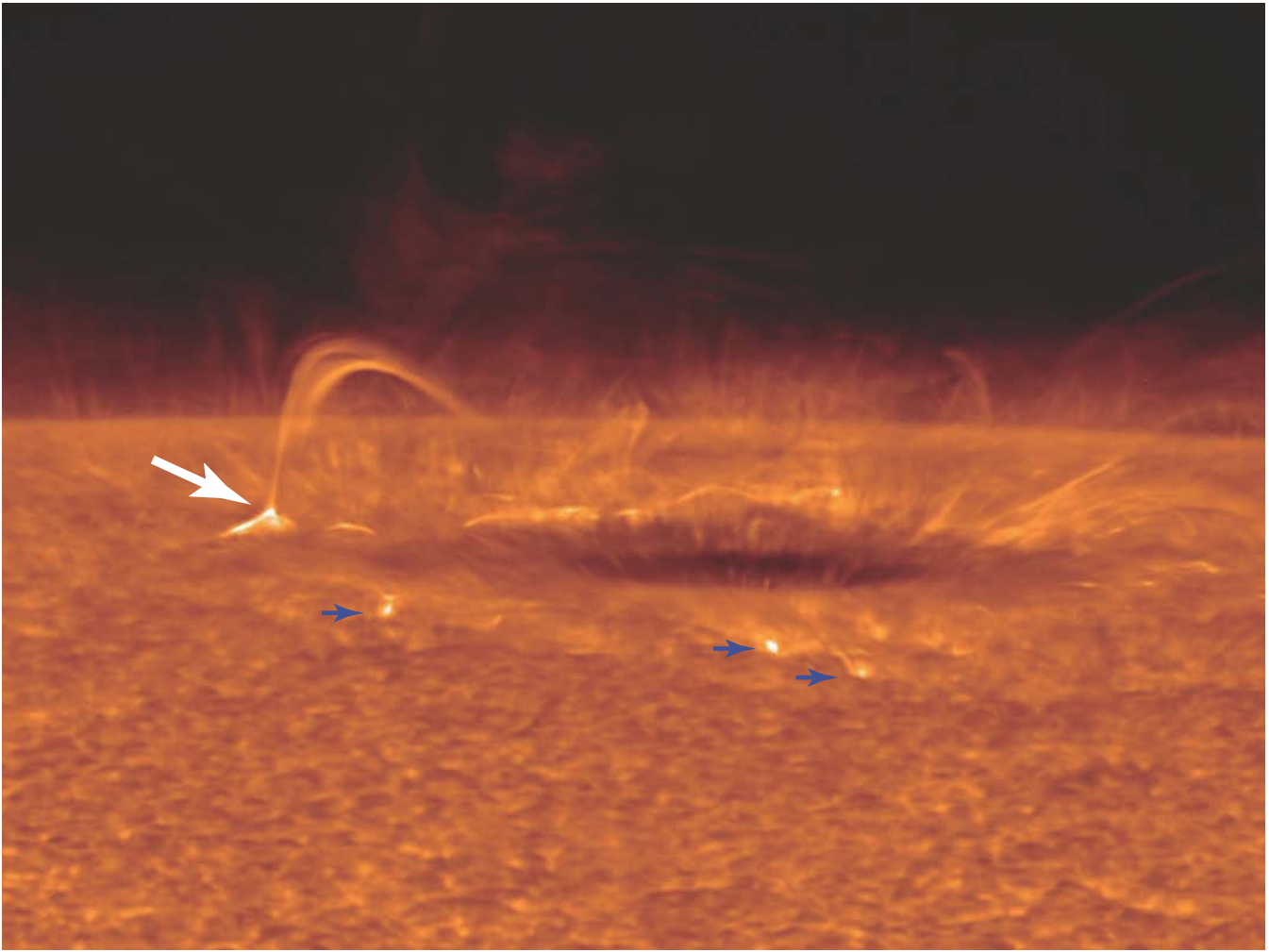

FIG. 5. A snapshot image of an active region recorded by using the Ca II H broadband filter. The image was processed to enhance small-scale anemone jets outside the sunspot (blue arrows) and a large-scale ejection (white arrow). The large-scale ejection has an inverse Y-shape, which is expected from the emerging flux model configuration. The many miniature rod-shaped features visible around the dark umbra are penumbral microjets. Courtesy of Y. Katsukawa.

### *2. Ellerman bombs*

Ellerman bombs are tiny, abrupt, short-lived brightness enhancements seen at the outer wings of chromospheric spectral lines, in particular the Hα line at 656.3 nm. The brightness enhancements at the outer wings are sometimes recognized as bi-directional Doppler signature, and they are not seen in the core of Hα because of shading by high-opacity overlying chromospheric fibrils. The bombs are observed in magnetic networks and their size is of the order of or under 1000 km. It is likely that some Ellerman bombs are observed at the footpoints of the chromospheric anemone jets[27], and they may underlie larger ejections, i.e., surges. Observations strongly suggest deep-seated photospheric reconnection of emergent or moat-driven magnetic flux with pre-existing strong vertical network fields as the mechanism underlying the Ellerman bombs[28,29,30]. In emerging flux regions where many Ellerman bombs are observed, many bombs are co-localized with neutral lines where the field lines could present an inverse Ω-shaped (or bald patch) configuration[31]. Multiple emerging bipoles may create an inverse Ω-shaped magnetic field configuration, and the converging motions at the photospheric level may trigger the reconnection[32 33 34].

### *3. Sunspot light bridges ejections*

Sunspot light bridges sometimes exhibit dynamical behaviors, including plasma ejections and simultaneous brightening events, which are well-observed in the Ca II H imaging observations. The sunspot light bridges are long bright lanes appearing in the sunspot dark umbra, dividing the umbra into two regions of same magnetic polarity. Recent observations indicate that weak-field gas penetrates by convection from below the photosphere into a vertically oriented strong magnetic field region, forming a magnetic field canopy structure in the photosphere. Inside the light bridges, the magnetic field has lower strength, is sparser, and is more horizontal than in the neighboring umbrae[35].

An interesting dynamics observed in some light bridges is long-lasting chromospheric plasma ejections[36]. Chromospheric plasma ejections were observed to be intermittently and recurrently produced for a long period[37]. The observed apparent ejected structures were only 1,500-3,000 km long (Figure 6c) when they were observed near the disk center of the Sun, but the actual length was shortened due to the projection effect. Assuming that the material is ejected along the direction extrapolated from the inclination of the measured vector magnetic field lines in the photosphere, the actual length of the ejected structures is estimated as 6,500–13,000 km. The apparent upward speed of ejectors is 6-40 km/s, derived from positional changes in the head of ejectors in successive series of Ca II H images. Considering the inclination of the magnetic field lines, the actual upward speed may be 26–180 km/s, which is supersonic and close to the Alfven speed in the corona. At the same time, the downward Doppler shift patch has also been detected at the ejectors footpoint locations (Figure 6f). This can be interpreted as a bidirectional jet wherein the down-flow patch detected in the photosphere is the downward component[38], while the chromospheric ejection seen in the Ca II H images is the upward component. We expect the reduced downward flow speed because density is high in the photosphere. The bidirectional jet is one of the features suggesting that magnetic reconnections occur at very low altitudes in the solar atmosphere.

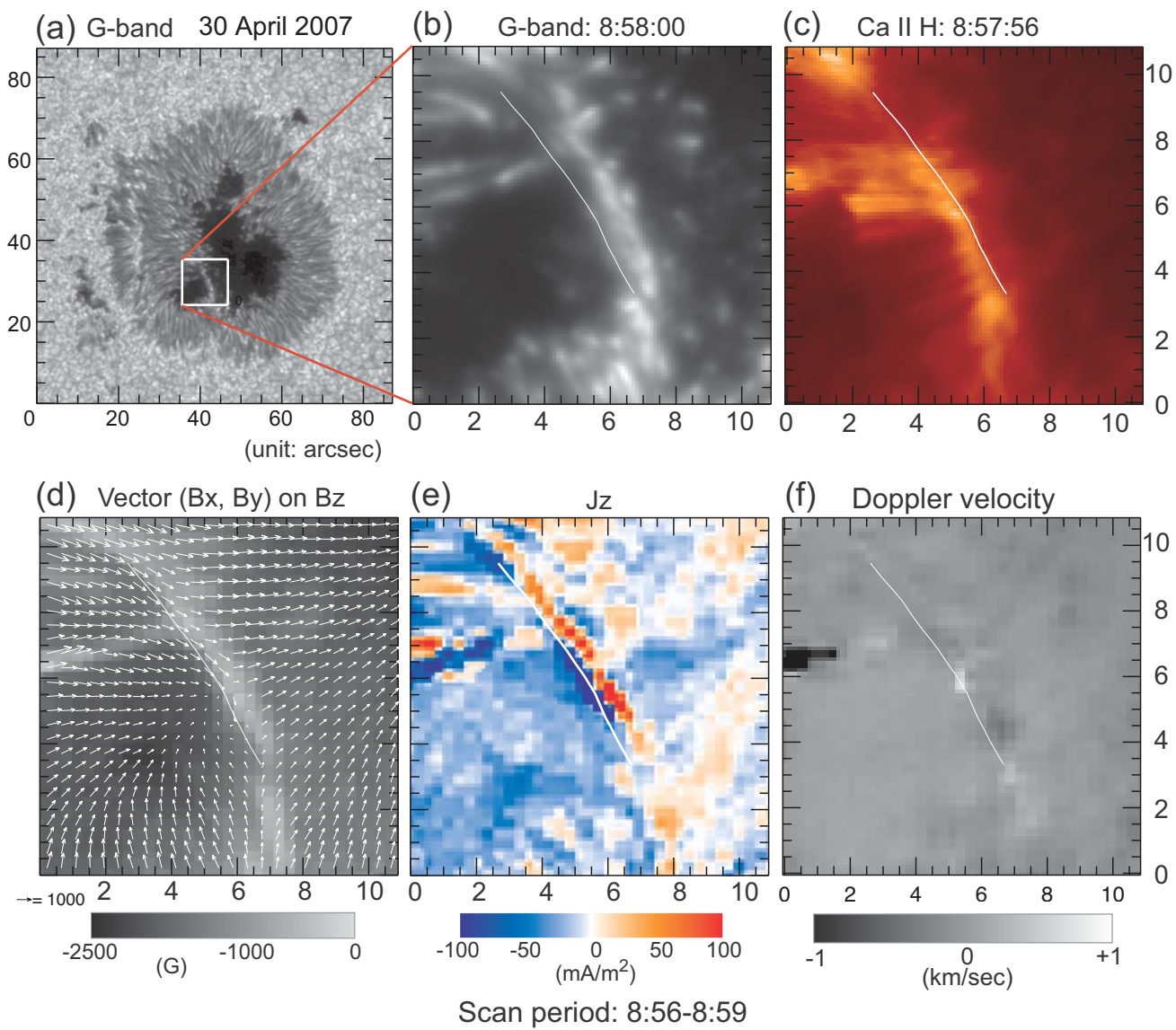

FIG. 6. A Ca II H intensity image showing an example of chromospheric ejections from sunspot light bridges and the corresponding photospheric maps. From (b) to (f): photospheric G-band intensity, chromospheric Ca II intensity, magnetic field vectors, vertical electrical current density ($J_z$), and Doppler shift in the photosphere. On the field vector map, the arrows show the horizontal components ($B_x$, $B_y$) of the magnetic flux density with gray scale for the vertical component ($B_z$). The solid lines in (b)-(f) denote the location of the right edge of ejection footpoints. The field of view is 14 arcsec × 14 arcsec and north is upward.

The photospheric magnetic vector field measurements may reveal obliquely oriented magnetic fields with strong vertical electric currents along the light bridge, as shown in Figure 6e. The current density $J_z$ is computed from the measured transversal field vector ($B_x$, $B_y$): $\mu_0 J_z = (\nabla \times B)_z = \partial B_y/\partial x - \partial B_x/\partial y$, where $\mu_0$ is the magnetic permeability. The strong electric current is located almost in the middle, not at the edge, of the light bridge. When the observed strong electric current is assumed to be a field-aligned current, the current flows along the magnetic field lines along the light bridge, yielding a helical magnetic flux tube along the light bridge, as illustrated in Figure 7. The current-carrying highly twisted magnetic flux tube is trapped along the light bridge below a canopy structure of the umbral fields. The poloidal component of the twisted flux tube is directed upward at the east side of the tube, resulting in the antiparallel configuration in the $Bz$ direction between the poloidal field and the neighboring vertically oriented umbral field. Antiparallel magnetic configurations are preferable for causing magnetic reconnection. Chromospheric ejections are dominantly observed at the east edge of the light bridge (Figure 5c).

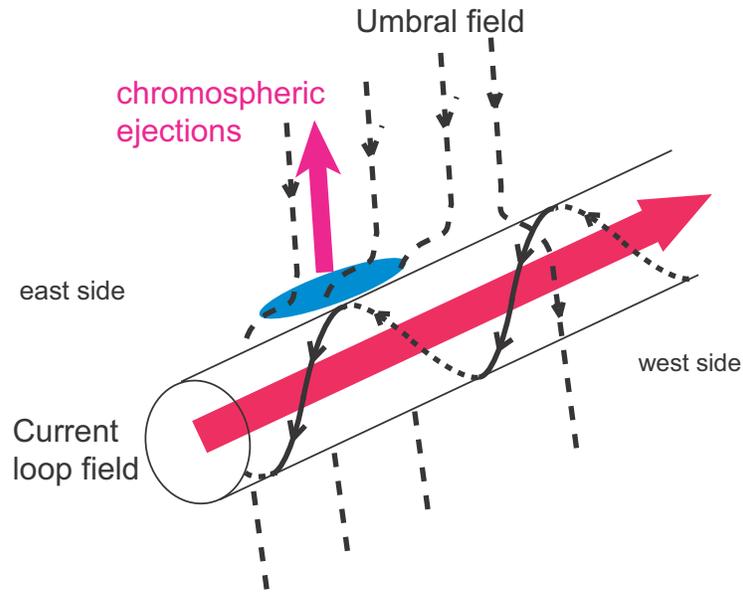

FIG. 7. A magnetic configuration proposed for sunspot light bridge ejections.

### *4. Sunspot penumbral microjets*

The Ca II H high cadence imaging observations have revealed the existence of tiny jet-like phenomena, referred to as penumbral microjets, in the chromosphere of sunspots penumbrae[39]. These events are not clearly seen in the non-processed Ca II H images, but they are clearly visible in the movies after high-pass filtering. They are more clearly seen in the sunspots located far from the disk center (Figure 5). The elongated collimated features are typically 1,000-4,000 km long and their width is at or under 400 km. Their lifetime is less than 1 min; they begin to brighten suddenly and darken gradually in several tens of seconds. Although it is difficult to clearly trace their temporal evolution in the elongated features, it appears that the brightening starts from the root of microjets, providing an apparent rise velocity of the upward plasma motion above 100 km/s at the start of the phenomena. The velocity is much faster than the acoustic velocity in the chromosphere. The energy was roughly estimated to be $2\times10^{23}$ ergs, which is of the order of nanoflares.

Penumbrae are the gray structures surrounding the dark umbrae of sunspots. Strong fields, above 2000 G, are observed in umbrae, and their orientation is generally close to the normal to the solar surface; The field lines are inclined by 40-50° at the umbra-penumbra boundary and gradually change into the inclined field lines (70-80° relative to the vertical) at the outer boundary of penumbra[40]. The penumbral magnetic field is non-homogeneous in the azimuthal direction on small scales. The continuum images reveal bright and dark filaments in penumbrae (Figure 3c). Magnetic field measurements suggest that the penumbral fine-scale field is composed of two magnetic components that differ in inclination: a flux-tube component and a more inclined field[41], as illustrated in Figure 8. The flux-tube component consists of flux tubes oriented more or less horizontally on the solar surface and the inclined field has an inclination of 40-50°. High-speed gas flows, referred to as the Evershed flows, may exist in the horizontal flux tubes. Due to the insufficient spatial resolution, it remains unclear whether helical fields exist in the flux tubes.

The penumbral microjets are likely launched from in between two horizontal flux tubes[39]. The jets direction may be aligned with respect to the inclined field between them. This interlaced magnetic configuration has the angle difference of about 20° between the horizontal and inclined fields, which possibly induces magnetic reconnection between them (Figure 8). Plasma flow measurements revealed a small (350 km or smaller in size) patch with downflow (about 1 km/s) in the photospheric layer, associated with brightening events seen in the Ca II H images, for some downflow patches[42]. The downflows may be a possible observational signature of downward outflow from magnetic reconnection and the outflow in the counter direction causes brightening in the chromosphere. Reproducing the microjets in computer simulations is challenging, with reconnecting field lines not being perfectly anti-parallel. A recent 3D numerical simulation[43] yielded a gas upflow, generated by the gas pressure gradient along the reconnected field lines; the flow tended to be directed parallel to the stronger, inclined magnetic field line when the field strength of the reconnecting field lines was asymmetric.

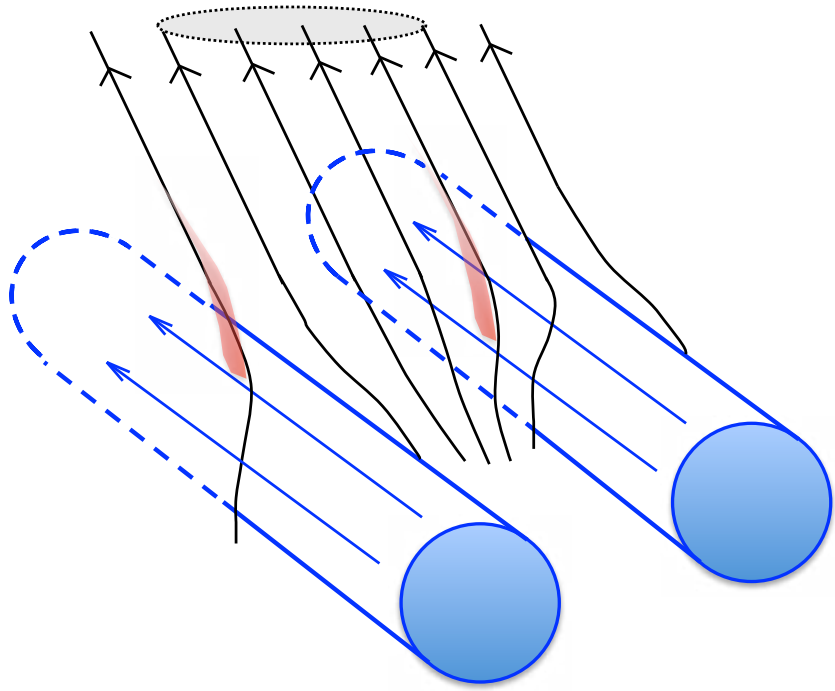

FIG. 8. The local fine-scale structure of the magnetic field in sunspot penumbrae and the launch of penumbral microjets from in between two horizontal flux tubes.

### *5. Spicules, RBEs, and rotating network field*

Spicules are dynamic, jet-like features in the chromosphere and are clearly observed in the Ca II movies of the solar limb (Figure 9). They are the fundamental structures connecting the solar photosphere to the hot corona and are believed to mediate energy and mass transfer for heating the corona. The traditional view is that the spicules' timescale is 3-15 min, including the rise-and-fall motions with the initial injection speed of 30-40 km/s[44]. In addition, recent *Hinode* Ca II observations revealed an extreme class of spicules, sometimes referred to as type II spicules[45]. They are much more dynamic with shorter lifetimes of 50-150 s and faster speed (30-150 km/s). They reach the height of 8,000-10,000 km[46] and thermally evolve to higher temperatures, resulting in the fading of chromospheric spicules[47]. Many, if not all, type II spicules exhibit swaying motion of the order of 15-20 km/s[48] and also twist or torsional motions of the order of 25-30 km/s on sub-arc second scales[49]. This motion may represent outward-propagating Alfven waves and is associated with rapid heating to at least TR temperatures[50]. Magnetic reconnection at the base of spicules is considered to drive such quick spicules, although magneto-acoustic shock waves may be the most preferable drivers of traditional spicules. There is a recent work claiming that there is no convincing example of type II spicules and that the majority of observed spicules are traditional spicules[51].

Most spicules originate from magnetic patches in magnetic networks, where the magnetic flux is unipolar[44]. On-disk features of spicules are observed as rapid blue-shift events (RBEs) in the chromospheric Hα line[52][53][54] and are possibly associated with UV blinkers in the transition region temperatures[21]. Some of them are likely observed as network small-scale jets on the solar disk in the UV lines originating from the transition region temperatures[55]. Identifying spicules on the solar disk can help to investigate the detailed magnetic properties at the spicules' roots. Because of unipolar magnetic fields in network regions, rotating motion of magnetic structures would be required for creating the reconnecting magnetic field components. Long-term continuous measurements of magnetic fields in the photosphere revealed rotating network magnetic field (RNF) in the quiet Sun[56]. When the magnetic fields rotate in the photosphere, rotation structures, referred to as cyclones, may be observed in the upper atmosphere[57]. The cyclones often accompany EUV brightening events, indicating the release of energy.

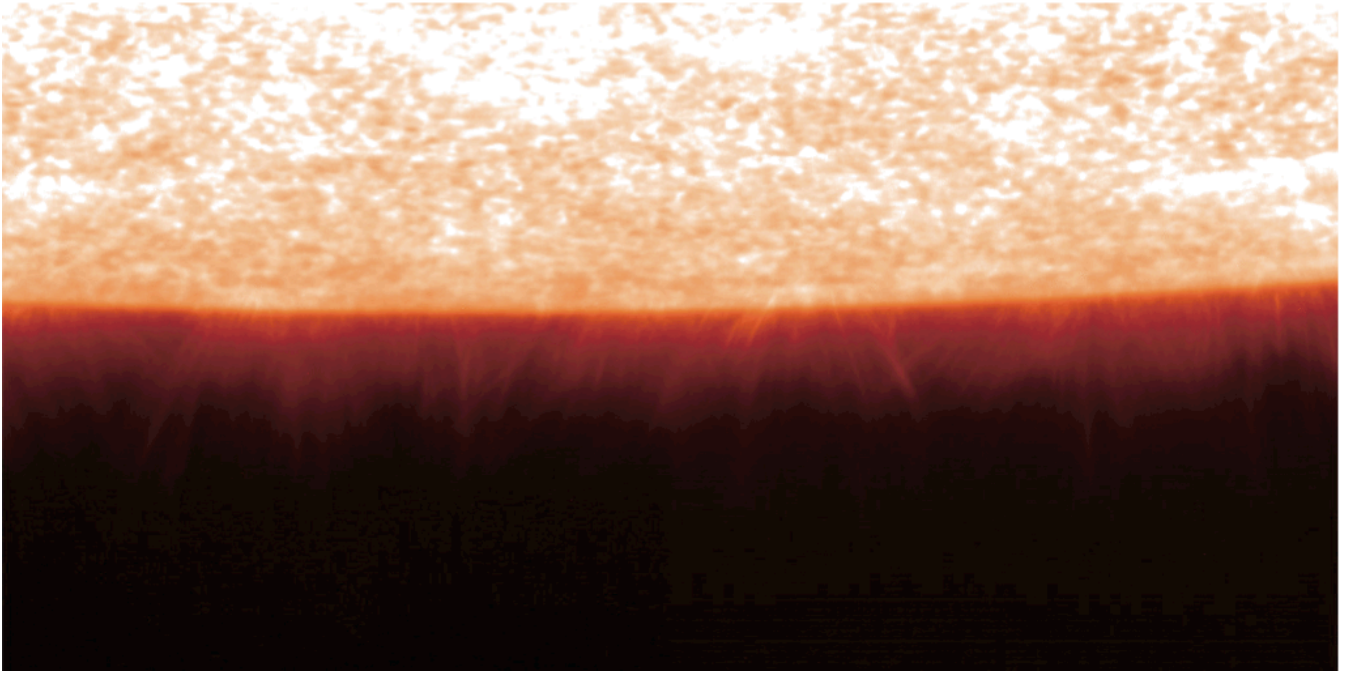

FIG. 9. Spicules at the south limb of the Sun, recorded by using the Ca II H broadband filter by the *Hinode* SOT.

## III. ANTI-PARALLEL RECONNECTION AND 3D COMPONENT RECONNECTION

Above, we briefly reviewed the properties of small-scale reconnection events with emphasis on the magnetic field configurations involved in the different events. In all of these events, magnetic reconnection has been suggested as the primary mechanism of jet-like ejections and transient heating. The observations suggest classification into two types of magnetic configurations: anti-parallel reconnection and 3D component reconnection.

**A. Anti-parallel reconnection**

In the anti-parallel reconnection, two anti-parallel magnetic field lines reconnect, which can be generally shown in 2D drawings. The formation of anti-parallel magnetic configuration is associated with either an emerging magnetic field or a moving magnetic field driven by the gas flow on the solar surface. The Ω type magnetic flux, emerging from below the solar surface, expands the emerging flux bundle above the solar surface because the gas pressure quickly decreases as a function of height. As shown in Figure 2a, the flux expansion may form a field discontinuity with the pre-existing nearby magnetic flux. The current sheet is enhanced at the field discontinuity, finally releasing the magnetic energy via magnetic reconnection. The moving magnetic flux driven by the surface gas flow can be another source for forming the field discontinuity. In the moat around the sunspot, the gas moves outward in the radial direction at an average speed of 1 km/s. Numerous moving flux patches, referred to as moving magnetic features (MMFs), are associated with the radial moat flow[58]. When opposite polarity patches exist at the destination of the moving flux, the field discontinuity is formed between the moving field and the opposite-polarity flux. Observations suggest that a small-scale emerging flux or a moving flux are found in about half of transient brightening events observed in active regions (active region transient brightenings or soft X-ray microflares); thus, the anti-parallel configuration has been proposed as the magnetic configuration. Magnetohydrodynamics simulations show that X-ray coronal jets and chromospheric ejections can be generated by magnetic reconnections in this magnetic configuration. It appears that the emerging flux can generate more intense transient brightening events than the moving magnetic flux.

In quiet regions, small-scale chromospheric jets are ubiquitously observed, as shown in Figure 5. The magnetic configuration proposed for chromospheric jets is the same as the emerging flux configuration proposed for X-ray coronal jets. The only difference between the two configurations is the size of the corresponding magnetic system, with smaller-size configurations resulting in jets that are observed in the chromosphere, not in the corona. Observations also suggest that the Ellerman bombs result from the reconnections of emergent or moat-driven magnetic fluxes with pre-existing vertical fields. These reconnections occur deeper in the solar atmosphere, i.e., the photosphere.

In general, the anti-parallel configuration is described in the 2D drawings, but the addition of a third component to the magnetic field configuration is needed for explaining some of the observed properties, such as untwisting evolution of ejectors. The third component may be called the guide field, which has been extensively discussed as one of key elements for realizing fast reconnections.

**B. 3D component reconnection**

The rest of the small-scale events reviewed in this article, i.e., multiple-loop-type soft X-ray microflares, sunspot light bridge ejections, sunspot penumbral microjets, and spicules, definitely require the 3D magnetic field configuration for magnetic reconnection. The reconnecting magnetic field lines are in the same direction, with a small shear angle between the field lines. In sunspot light bridge ejections, one magnetic flux tube is embedded in the vertically oriented field (Figure 7). The flux tube may be helical or twisted and the shear angle between the helical field and the nearby field is only 40-50°[37]. Sunspot penumbral microjets are launched from the magnetic discontinuity, where the magnetic shear angle is about 20° (Figure 8). There might be a smaller shear angle at the source of chromospheric spicules, because of the unipolar region. An important question is the condition for the onset of this reconnection for generating these small-scale events in the solar atmosphere. Many field discontinuities with small shear angles exist in the solar atmosphere[59]. However, these small-scale events are localized in magnetic networks and sunspots, suggesting that small-scale events may occur only for specific magnetic field discontinuities formed in the solar atmosphere. One hint for considering this suggestion may be provided by recent magnetopause observations, which have demonstrated clear dependence of magnetic reconnection on the plasma $\beta$ and magnetic shear angle[60]. When the difference $\Delta\beta$ between the plasma $\beta$ values on both sides of the field discontinuity is large, magnetic reconnection events are observed only with large magnetic shear. For low $\Delta\beta$, reconnections occur for both low and high magnetic shears. Particle-in-cell (PIC) simulations of reconnections have demonstrated that diamagnetic drifts in plasma with high $\beta$ but small shear angles suppress reconnections[61]. Large $\Delta\beta$ is expected to exist in the chromosphere because the complicated atmospheric structure exhibits significant density variations and magnetic field intensity variations in the chromosphere. For example, at the boundary of sunspot light bridges, where light bridge ejections are typically launched from, there is a large difference between magnetic field intensities in and out of light bridges, suggesting a large $\Delta\beta$. When the magnetic shear angle between the light bridge field and sunspot vertical field does not exceed 40-50°, reconnections may be suppressed by diamagnetic suppression; reconnections may be triggered when the shear angle exceeds 50°.

To investigate this possibility, 3D numerical simulations and laboratory plasma collision experiments would be helpful. Recently, we experimentally reproduced the magnetic configuration of sunspot light bridge ejections in laboratory plasma, and investigated the dynamics of plasma ejections and heating triggered by component reconnection[62]. We succeeded in qualitatively reproducing ejections and heating, although there remain quantitative differences between observations and experiments, such as the differences between jet velocity and total energy values. This experiment used only one plasma condition, but we will be able to investigate reconnection onset conditions in more detail if different plasma conditions, i.e., different magnetic shear angles and plasma $\beta$ gradients $\Delta\beta$, can be created in the laboratory environment.

Two (or more) magnetic flux tubes are probably involved in the magnetic configuration for multiple-loop-type microflares. The simplest configuration consists of two parallel current-carrying helical loops, as shown in Figure 4. The main axis components of the magnetic fields in the helical loops do not reconnect, whereas the poloidal components of the helical loops are anti-parallel, resulting in reconnection. This configuration is probably the same as co-helicity merging configuration in laboratory plasma collision experiments[63]. Because the energy released by multiple loop brightening events is larger than that of single brightening[17], the magnetic reconnection of parallel or tilted twisted flux tubes may be more important for creating larger-scale energy release events in the corona. Quantitative comparisons between solar small-scale reconnection events and merging experiments can provide additional information.

## IV. SUMMARY

The solar atmosphere is replete with small-scale transient phenomena with brightening events and jet-like ejections. These events occur at different locations on the solar surface and vary in their morphology and amount of the released energy. Observations suggest that these magnetic configurations can be classified into two groups. The first group is anti-parallel configurations, represented by the standard emerging flux configuration. The other group is 3D component configurations, in which reconnections among parallel or tilted twisted flux tubes occur. The latter may be more important for the larger class of small-scale reconnection events, although further qualitative studies are needed. The two types of magnetic configurations can be compared to counter-helicity and co-helicity merging configurations, respectively, in laboratory plasma collision experiments.

## ACKNOWLEDGMENTS

This article is based on the talk presented by T.S. at the US-Japan MR2014 workshop (June 2014) in Tokyo, Japan. T.S. wishes to thank Prof. Y. Ono for encouraging him to write this article. *Hinode* is a Japanese mission developed and launched by ISAS/JAXA, with NAOJ as domestic partner and NASA and STFC (UK) as international partners. It is operated by these agencies in co-operation with ESA and NSC (Norway). This work was partially supported by JSPS KAKENHI Grant Number 23540278, 15H05750 and JSPS Core-to-Core Program 22001.